# Numerical modeling of CuSbSe$_2$-based dual-heterojunction thin film solar cell with CGS back surface layer

Bipin Saha[1], Bipanko Kumar Mondal[1,2], Shaikh Khaled Mostaque[1], Mainul Hossain[3] and Jaker Hossain[1*]

[1]*Solar Energy Laboratory, Department of Electrical and Electronic Engineering, University of Rajshahi, Rajshahi 6205, Bangladesh.*

[2]*Department of Electrical and Electronic Engineering, Pundra University of Science & Technology, Bogura, Bogura 5800, Bangladesh.*

[3]*Department of Electrical and Electronic Engineering, University of Dhaka, Dhaka-1000, Bangladesh.*

Ternary chalcostibite copper antimony selenide (CuSbSe$_2$) is a promising absorber material for next generation thin film solar cells due to the non-toxic nature, earth-abundance, low-cost fabrication technique, optimum bandgap and high optical absorption coefficient of CuSbSe$_2$. Conventional single heterojunction CuSbSe$_2$ solar cells suffer from high recombination rate at the interfaces and the presence of a Schottky barrier at the back contact, which limit their power conversion efficiencies (PCEs). In this study, we propose a dual-heterojunction $n$-ZnSe/$p$-CuSbSe$_2$/$p^+$-CGS solar cell, having copper gallium selenide (CGS) as the back surface field (BSF) layer. The BSF layer absorbs longer wavelength photons through a tail-states-assisted (TSA) two-step upconversion process, leading to enhanced conversion efficiency. Numerical simulations were carried out using SCAPS-1D to investigate the performance of the proposed solar cell with respect to absorber layer thickness, doping concentrations and defect densities. The simulation results exhibit PCE as high as 43.77% for the dual-heterojunction solar cell as compared to 27.74% for the single heterojunction *n*-ZnSe/*p*-CuSbSe$_2$ counterpart. The dual-heterojunction structure has, therefore, the potential to approach the Shockley-Queisser (SQ) detailed balance limit and can lead to extremely high PCEs in emerging thin film solar cells.

## 1. Introduction

Solar energy provides a safer and cleaner alternative to the rapidly depleting fossil fuels. At present, the photovoltaic (PV) industry is dominated by homojunction crystalline-silicon (c-Si) solar cells. Silicon is the material of choice because it is earth-abundant, non-toxic and highly stable. However, single junction silicon PV devices are unable to surpass the theoretical Shockley-Queisser (S-Q) efficiency limit of 32.33% [1,2]. Recent studies have, therefore, focused on heterojunction solar cell structures, with PCEs approaching the S-Q limit. Yoshikawa *et al.* [3], for example, experimentally demonstrated an interdigitated back contact (IBC) Si heterojunction solar cell with a power conversion efficiency (PCE) of 26.6% and a realistic module size of 180 cm$^2$. Despite the success, silicon-based solar cells are handicapped by the low absorption coefficient in longer wavelengths and high processing temperature of Si (~1400 °C) [4]. Moreover, defect levels in amorphous-Si (a-Si), limits the further developments of a-Si/c-Si heterojunctions [5]. Although, cadmium telluride (CdTe) and copper indium gallium arsenide (CIGS) thin films exhibit good absorption in the visible solar wavelength and provide modest power conversions, their efficiencies are still inferior to that of Si. Apart from the toxic nature of these materials, the manufacturing costs are increasingly dominated by constituent materials like the top cover sheet and other encapsulants [6,7]. Third-generation solar cells, based on III-V materials and their tandems are also being investigated, as a viable alternative to Si, for achieving high PCE solar cells [8]. Many studies have concentrated their efforts on perovskite [9–11] and polymer [12] solar cells. However, despite the high PCEs, both perovskite and polymer solar cells suffer from stability issues, which limit their long-term applicability [13]. Highly stable, inorganic, dual-heterojunction solar cells promise to deliver high PCE at low cost. Our previous simulations have shown CdTe-based dual-heterojunction solar cells, with a back surface-field layer, achieving a PCE over 40% [14].

Here, we propose a dual-heterojunction $n$-ZnSe/$p$-CuSbSe$_2$/$p^+$-CGS solar cell, having ZnSe as the window layer and CGS (CuGaSe$_2$) the back surface field (BSF) layers. The $p$-doped CuSbSe$_2$ serves as the main absorber layer with a carrier concentration of $5.6 \times 10^{15}$ cm$^{-3}$, tunable energy bandgap (1.0-1.6 eV) and high absorption coefficient (~$10^4$ cm$^{-1}$). CuSbSe$_2$ is a promising absorber material for thin film solar cells, due to its excellent electrical and optical properties, low cost, earth abundance and non-toxic nature [15]. Unlike Si, which requires a very high processing temperature, the growth of CuSbSe$_2$ is favored at low temperatures ranging between 380-410°C [16,17]. The significant absorption of light in CuSeSb$_2$ is due to the presence of the lone pair of 5s$^2$ electrons. CuSbSe$_2$ reduces the surface roughness and back surface recombination in the absorption layer, facilitating the collection of carriers [18]. This can be explained in terms of the orthorhombic structure of CuSbSe$_2$, in which the lone pair electron configuration of Sb distorts the tetragonal bonds, resulting in a layered structure with zero dangling bonds which are primarily responsible for grain boundary carrier recombination. Hence, CuSbSe$_2$ structure is more defect tolerant [19]. ZnSe is chosen as the window layer, because of its wide bandgap (2.7 eV), earth-abundance and high optical transmission in the visible range [20–23]. The thin layer of ZnSe allows more than 80% of the incoming light to reach the underlying CuSbSe$_2$ absorber layer without being absorbed [20,24,25]. In addition, as compared to other more commonly used window layers, ZnSe is less sensitive to moisture and oxidation [21]. CGS (CuGaSe$_2$) has an energy band gap of 1.68 eV and a high absorption coefficient in longer wavelengths [26,27], which make CGS an excellent BSF layer for enhancing the PCE of *n*-ZnSe/*p*-CuSbSe$_2$/*p$^+$*-CGS solar cell.

In this work, we use numerical simulations to evaluate the performance of the proposed dual-heterojunction *n*-ZnSe/*p*-CuSbSe$_2$/*p$^+$*-CGS solar cell, with respect to thickness, doping concentrations and defect levels of the absorber, window and BSF layers. All device simulations are carried out using Solar Cell Capacitance Simulator-1 dimensional (SACPS-1D) software (SCAPS-1D), which solves Poisson's equation and the continuity equations for free electrons and holes [28].

## 2. Device structure and simulation model

Figures 1(a) and (b) show the schematic and the energy band diagram of the simulated dual-heterojunction *n*-ZnSe/*p*-CuSbSe$_2$/*p$^+$*-CGS, respectively. In the structure, the incident sunlight passes through the ZnSe layer and is absorbed in the CuSbSe$_2$ and CGS layers, respectively. The electron affinity and ionization potential of ZnSe are 4.09 and 6.79 eV, respectively [24] whereas those of CuSbSe$_2$ are 4.11 and 5.19 eV, respectively [29]. Therefore, ZnSe forms a suitable pn heterojunction with CuSbSe$_2$ compound. On the other hand, CGS has an electron affinity and band gap of 3.61 and 1.66 eV, respectively. As a result, it can also form a suitable $pp^+$ heterojunction with CuSbSe$_2$ layer. The electron and hole quasi-Fermi levels are $E_{Fn}$ and $E_{Fp}$, respectively as shown in Fig. 1(b). The electrons generated in the CuSbSe$_2$ and CGS layers can easily transport to the cathode and holes generated in the CuSbSe$_2$ and CGS can also move towards the anode favored by the suitable energy barriers.

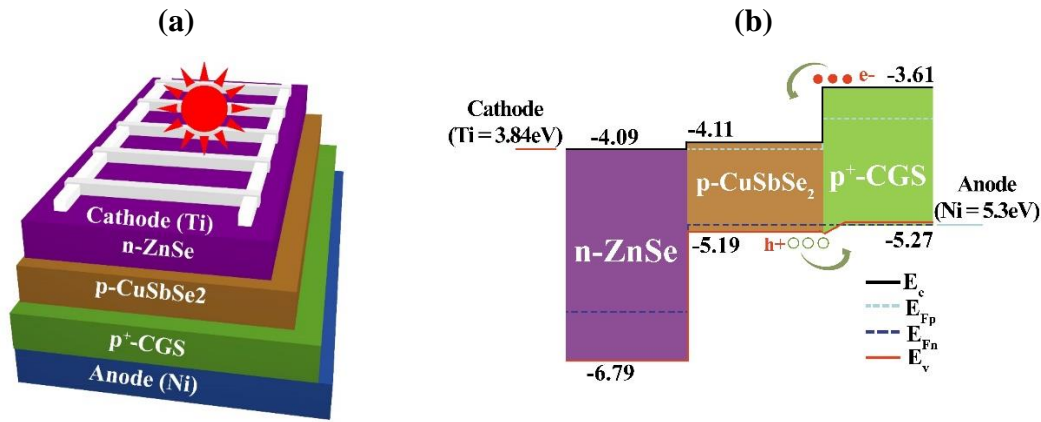

**Fig. 1:** (a) The simulated structure and (b) illuminated energy band diagram of the modeled $n$-ZnSe/$p$-CuSbSe$_2$/$p^+$-CGS dual-heterojunction solar cell.

The numerical simulations were performed under the illumination of one sun with 100 mW/cm$^2$ and a global air mass (AM) of 1.5G spectrum. The operating temperature was taking as 300 K. The ideal values of series and shunt resistance were used and radiative recombination coefficient was not considered. Acceptor and donor doping profiles in the bulk layers were considered to be Gaussian and interface defects were also taken into account. The experimental values of absorption coefficient for ZnSe, CuSbSe$_2$ and CGS layer were obtained from previous studies [15,33]. SCAPS-1D automatically takes into consideration, the sub-band gap absorption effect from the optical data. The simulation parameters used for different layers are provided in Table 1, and interfaces parameters are provided in Table 2.

**Table 1.** Input parameters used for simulating ZnSe/CuSbSe$_2$/CGS heterojunction solar cells at 300 K.

| Parameters | $n$-ZnSe [30] | $p$-CuSbSe$_2$ [31] | $p^+$-CGS [32] |
|---|---|---|---|
| Layer | Window | Absorber | BSF |
| Conductivity | N-type | P-type | P$^+$-type |
| [a] Thickness [μm] | 0.2 | 0.8 | 0.2 |
| Bandgap E$_G$ [eV] | 2.7 | 1.080 | 1.660 |
| Electron affinity χ [eV] | 4.09 | 4.110 | 3.610 |
| Effective Density of CB, N$_C$ [cm$^{-3}$] | 1.5×10$^{18}$ | 9.9×10$^{19}$ | 2.2×10$^{17}$ |
| Effective Density of VB, N$_V$ [cm$^{-3}$] | 1.8×10$^{19}$ | 9.9×10$^{19}$ | 1.8×10$^{18}$ |
| Electron Mobility μ$_n$ [cm$^2$V$^{-1}$s$^{-1}$] | 50 | 10 | 100 |
| Hole mobility μ$_p$ [cm$^2$V$^{-1}$s$^{-1}$] | 20 | 10 | 250 |
| [a] Donor Concentration N$_D$ [cm$^{-3}$] | 1×10$^{18}$ | 0 | 0 |
| [a] Acceptor Concentration N$_A$ [cm$^{-3}$] | 0 | 1×10$^{16}$ | 1×10$^{18}$ |
| Type of defect | Acceptor | Donor | Neutral |
| Energetic distribution | Gaussian | Gaussian | Single |
| [a] Peak defect density, N(t) [eV$^{-1}$cm$^{-3}$] | 5.642×10$^{13}$ | 5.642×10$^{13}$ | 1×10$^{13}$ |
| Characteristics energy [eV] | 0.1 | 0.1 | 0.1 |
| Reference Energy [eV] | 1.350 | 0.650 | 0.6 |
| Capture cross section of electron for acceptor defect [cm$^2$] | 1×10$^{-15}$ | 1×10$^{-15}$ | 1×10$^{-15}$ |
| Capture cross section of hole for acceptor defect [cm$^2$] | 1×10$^{-15}$ | 1×10$^{-17}$ | 1×10$^{-15}$ |

[a] variable field.

**Table 2.** Interface input parameters used in simulation

| Parameters | CuSbSe$_2$/p$^+$-CGS | CuSbSe$_2$/n-ZnSe |
|---|---|---|
| Defect type | Neutral | Neutral |
| Capture cross section for electrons [cm$^2$] | $1\times10^{-19}$ | $1\times10^{-19}$ |
| Capture cross section for holes [cm$^2$] | $1\times10^{-19}$ | $1\times10^{-19}$ |
| Energetic distribution | Single | Single |
| Reference for defect energy level Et | Above the highest EV | Above the highest EV |
| Energy with respect to reference [eV] | 0.6 | 0.6 |
| Total defects [cm$^{-2}$] | $1\times10^{9}$ | $1\times10^{10}$ |

## 3. Results and discussion

### *3.1 CuSbSe$_2$ absorber layer*

Figure 2(a) illustrates the electrical characteristics of the proposed solar cell with varying CuSbSe$_2$ absorber layer thickness. The short circuit current ($J_{sc}$), open circuit voltage ($V_{oc}$), fill factor (*FF*) and overall PCE of photovoltaic cell were calculated for CuSbSe$_2$ thickness, ranging between 0.6 and 1.0 µm. A thicker absorber layer absorbs more photons and generates large number of electron-hole pairs. Hence, $J_{sc}$ increases from 57.94 mA/cm$^2$ when CuSbSe$_2$ is 0.6 µm thick, to 59.58 mA/cm$^2$ when the absorber layer thickness becomes 1 µm. In contrast, $V_{oc}$ the slightly decreased from 0.94 V to 0.92 V after increasing the absorber thickness from 0.6 µm to 1 µm. This is because of the increase in recombination current with increasing CuSbSe$_2$ thickness. The FF and PCE remains fairly constant, ranging between 79.72 and 79.25% for the FF and 43.76 and 43.64% for the

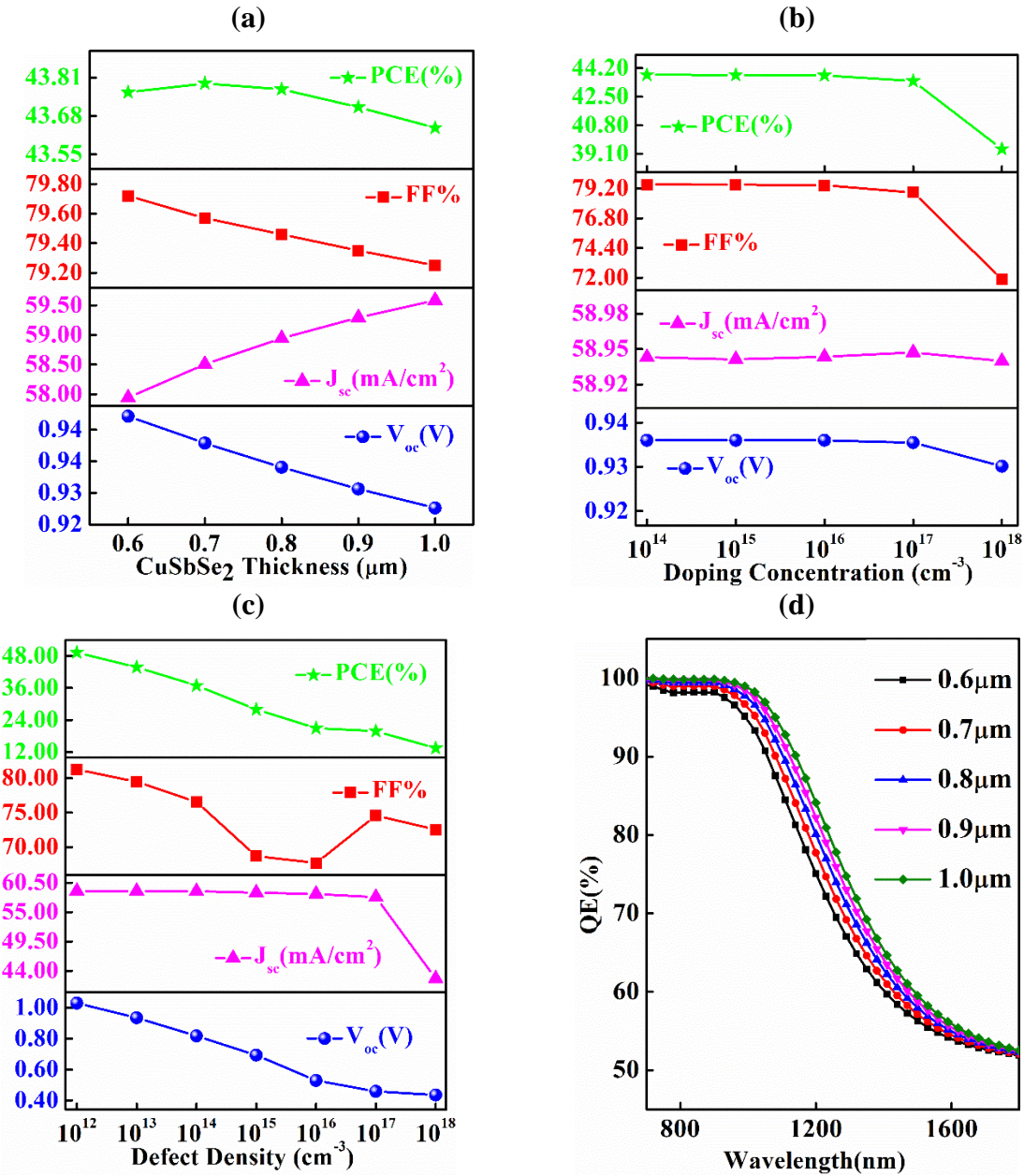

**Fig. 2.** Output PV parameters with respect to (a) thickness (b) doping concentration (c) bulk defects of the CuSbSe$_2$ absorber layer (d) The effect of CuSbSe$_2$ thickness on quantum efficiency (QE) of the solar cell device.

PCE. The maximum PCE is 43.79% and is obtained with a 0.7 μm thick CuSbSe$_2$ absorber layer. Figure 2(b) shows the change in device characteristics when the doping concentration in CuSbSe$_2$ is varied. $J_{sc}$ remains almost unchanged, varying only slightly, between 58.94

mA/cm$^2$ and 58.93 mA/cm$^2$ for doping concentrations of $10^{14}$ to $10^{18}$ cm$^{-3}$, respectively. $V_{oc}$ follows a similar trend, maintaining a steady value of around 0.93 V. The *FF*, however, drops from 78.88% to 71.86%, below a doping level of $10^{17}$ cm$^{-3}$, affecting the overall PCE. The PCE varies between 43.8% and 39.39%, between doping concentrations of $10^{14}$ to $10^{18}$ cm$^{-3}$, respectively.

The variation of PV parameters with CuSbSe$_2$ bulk defects is exhibited in Fig. 2(c). The bulk defect density ranges from $10^{12}$ to $10^{18}$ cm$^{-3}$. A high defect level impedes the generation of electron-hole pairs, leading to decreasing $J_{sc}$ with increasing defect density. Similarly, $V_{OC}$ also decreases abruptly from 1.03 to 0.44 V due to higher defects. The Shockley-Read-Hall (SRH) recombination is the major recombination mechanism at high defects, leading to an increase in the reverse saturation current and a decrease in $V_{OC}$. The *FF* drops from 81.26 to 72.55% due to the high defects of CuSbSe$_2$ layer, causing the corresponding PCEs to vary from 49.32 to 13.45 %.

Figure 2(d) shows how the quantum efficiency (QE) of the device changes with CuSbSe$_2$ thickness. As expected, thicker absorber layer results in higher QE due to higher number of absorbed photons. At shorter wavelengths, QE is almost independent of CuSbSe$_2$ thickness. However, at longer wavelength regime, QE decreases as photon energy hv of the incident light < band gap E$_g$ of CuSbSe$_2$.

*3.2 ZnSe window layer*

In this section, the effect of ZnSe window layer on the performance of *n*-ZnSe/*p*-CuSbSe$_2$/*p*$^+$-CGS dual-heterojunction solar cells has been studied. The change in PV parameters for varying thickness of ZnSe are shown in Fig. 3(a). The window layer thickness was varied from 0.1 to 0.5 μm, and the corresponding $J_{SC}$, $V_{OC}$, *FF* and *PCE* values are calculated. It is noticed in the figure that a modest increment in short circuit current $J_{sc}$, occurs when the layer thickness of the window ZnSe layer is increased. Over the thickness range, $V_{OC}$ *FF* and PCE also remain constant.

Fig. 3(b) illustrates the $J_{SC}$, $V_{OC}$, *FF* and *PCE* with changes in doping concentrations in the ZnSe layer. The doping level in ZnSe was altered from $10^{16}$ to $10^{20}$ cm$^{-3}$. Both $J_{SC}$ and $V_{OC}$ remained fairly unchanged, with $J_{SC}$ and $V_{OC}$ showing a slightly decreasing trend and e increasing trends, respectively, as the doping concentration is raised from $10^{16}$ to $10^{20}$ cm$^{-3}$. The FF and PCE follows the same trend with the PCE ranging from 43.63% to 43.79% for specified ZnSe doping range.

The defect density of ZnSe also affects the PV parameters as depicted in Fig. 3(c). Bulk defects of ZnSe window layer have negligible effects on the $J_{SC}$ and $V_{OC}$ but the *FF* and PCE decreases at defect levels, greater than $10^{17}$ cm$^{-3}$.

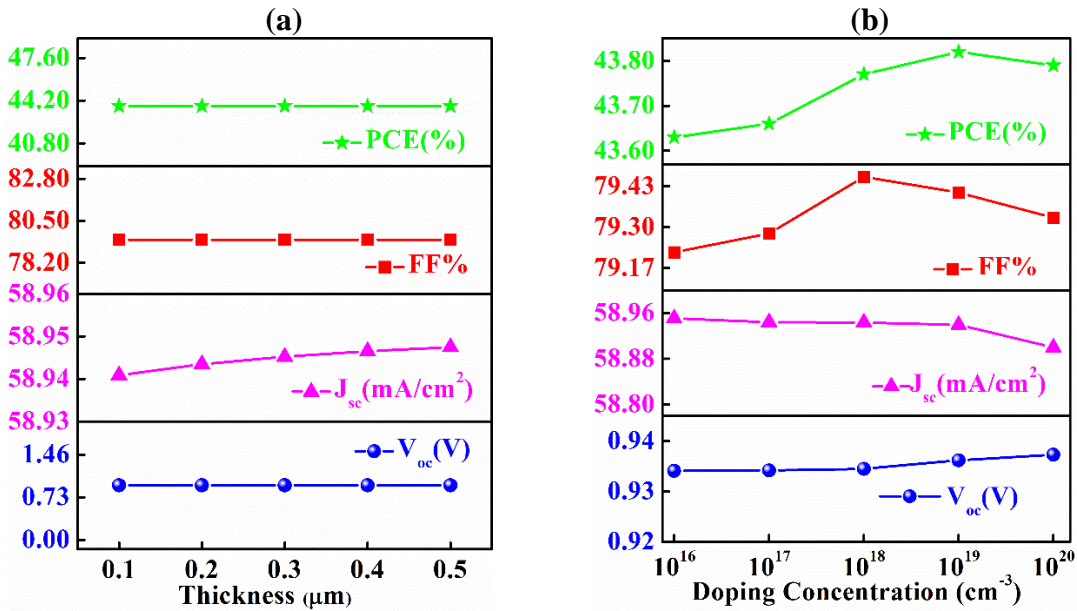
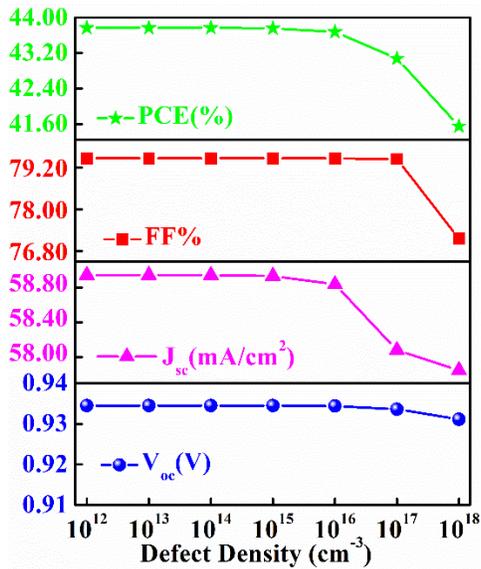

**Fig. 3**. Output PV parameters with respect to (a) thickness (b) doping concentration (c) bulk defects of the ZnSe window layer.

*3.3 CGS BSF layer*

The CGS BSF layer serves as the bottom absorber layer and contributes significantly to the performance of the proposed device. The PV parameters with respect to the thickness of CGS BSF layer is shown in Fig. 4(a). The layer thickness of the CGS layer was varied from 0.1 to 0.5 µm. $J_{SC}$ was found to be 44.07 mA/cm$^2$ without the BSF layer. But when only 0.1 µm thick CGS layer was employed in the structure, the current is significantly enhanced to 54.43 mA/cm$^2$. Furthermore, $J_{SC}$ continuously increased with increasing thickness of CGS layer. reaching 65.04 mA/cm$^2$, when the CGS became 0.5 µm thick. This occurs because CGS BSF layer absorbs longer wavelength incident photons through a two-step, tail-state-assisted (TSA) photon upconversion process [34,35]. In TSA two-step upconversion process, two low energy photons are absorbed in a series by Urbach tail states, creating an extra electron-hole pair [32,36]. This in turn, leads to significantly higher Jsc with increasing CGS BSF layer thickness. Correspondingly, open circuit voltage V$_{oc,}$ increases from 0.73 to 0.93 V with CGS thickness. The insertion of CGS layer generates high built-in potential in CuSbSe$_2$/CGS interface which is the responsible for high $V_{OC}$. The *FF* decreases when the CGS layer is introduced. However, it remains constant with the change in thickness of CGS layer. Since, both $J_{SC}$ and $V_{OC}$ are enhanced, the overall PCE increases from 40.21% to 48.61%, over the range of CGS thickness.

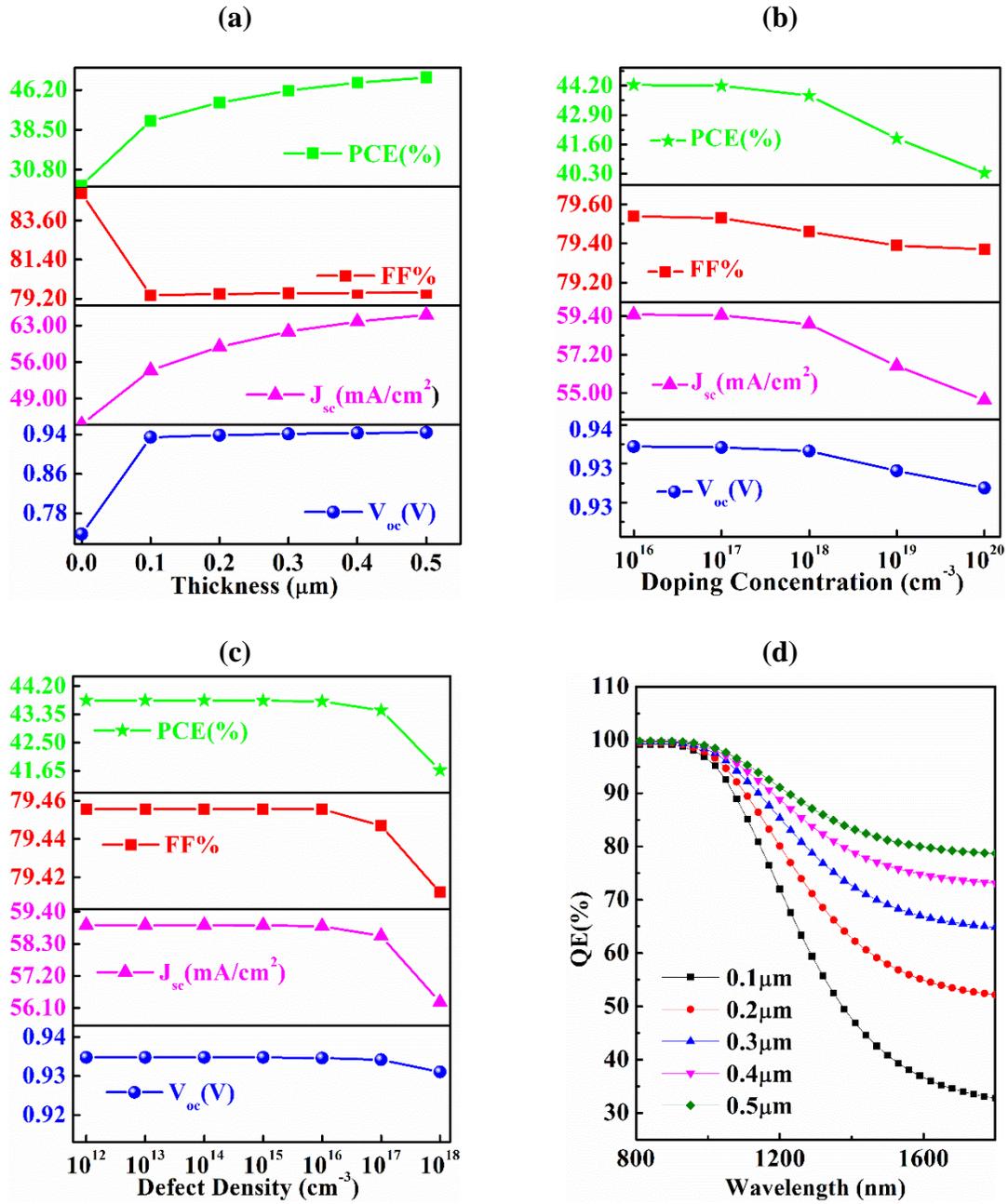

**Fig. 4.** PV parameters with respect to (a) thickness (b) doping concentration (c) bulk defects of the CGS BSF layer (d) The effect of CGS BSF layer thickness on quantum efficiency (QE) of the solar cell device

Figure 4(b) shows the effect of doping concentration of CGS BSF layer on the PV parameters. The carrier concentration of CGS layer has been adjusted from $10^{16}$ to $10^{20}$ cm$^{-3}$. $J_{SC}$ changes abruptly from 59.49 to 54.60 mA/cm$^2$ when the carrier concentration was increased beyond $10^{18}$ cm$^{-3}$. The decrement of short circuit current is reasonable because of parasitic free carriers which are absorbed at high carrier concentration. $V_{OC}$ slightly decreases from 0.935 V to 0.930V, while the FF remains unchanged with increasing in doping concentrations. Overall, the PCE drops from 44.24% to 40.33% with the increment of the carrier concentrations in the BSF.

The variation in PV parameters with bulk defects in the CuSbSe$_2$ BSF layer, as shown in Fig. 4(c), has also been studied. The presence of defects in the BSF layer inhibits free carrier generation. Higher level ($> 10^{17}$ cm$^{-3}$) of bulk defects in CuSbSe$_2$ causes the $J_{SC}$ to drop from 58.94 to 56.30 mA/cm$^2$ when the defect densities change from $10^{12}$ to $10^{18}$ cm$^{-3}$. The bulk defects, however, have negligible impact on $V_{OC}$ and $FF$. Finally, the PCE also decreases, from 43.77 to 41.67% with increase in defect density.

The change in QE of $n$-ZnSe/$p$-CuSbSe$_2$/$p^+$-CGS dual-heterojunction solar cells, with respect to CGS BSF layer thickness, is shown in Fig. 4(d). QE enhances gradually with the increment of CGS thickness. At any given thickness of CGS, in the longer wavelength region, QE remains significantly high. For example, at a wavelength of 1600 nm, the QE > 50% at with 0.2 μm thick CSG BSF layer. This confirms the absorption of longer wavelength photons, by the CGS layer, through TSA two-step photon upconversion, leading to a significant improvement in PCE.

## 3.4 Interface defects

Defect densities at the CuSbSe$_2$/ZnSe and CGS/CuSbSe$_2$ interfaces influence the solar cell performance as shown in Figures 5(a) and (b). High defect levels at the CuSbSe$_2$/ZnSe interface cause the $V_{OC}$ to drop from ~0.94 V to ~0.80 V, as depicted in Fig. 5(a). $J_{SC}$ remains fairly constant while the FF increases from 79.2% to 83.6%, causing the PCE to drop from 43.5% to 39% as defect density increases. A similar trend, as shown in Fig. 5(b), is observed for the CGS/CuSbSe$_2$ interface

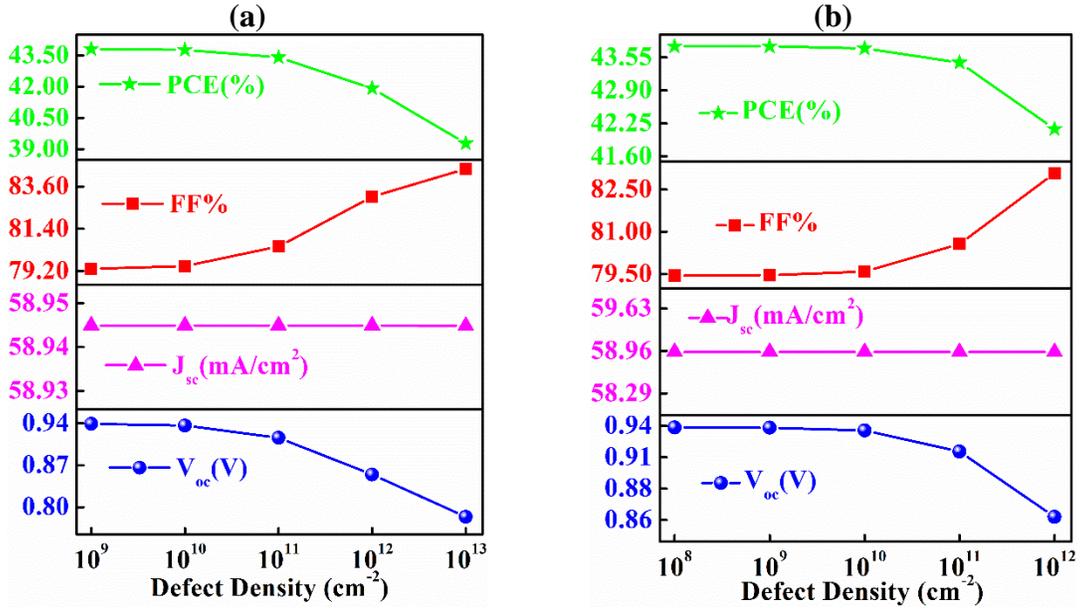

**Fig 5.** PV parameters with respect to interface defect density at (a) CuSbSe$_2$/ZnSe and (b) CGS/CuSbSe$_2$ interfaces.

## 3.5 Temperature dependence

Carrier mobilities in semiconductors decrease with increasing temperatures, leading to a reduction in bond energies. This, in turn, causes the bandgap to decrease, affecting the absorption properties. Because of the square relationship between reverse saturation current ($I_o$) and intrinsic carrier concentration ($n_i$) in solar cells, $V_{OC}$ is highly impacted by temperature fluctuations. At higher temperatures, the enhanced recombination rate of electrons and holes lowers the number of free carriers in the cell. Fig. 6 delineates how the increase in temperature from 273 to 500 K affects heterojunction solar cell performance. It is observed that the $J_{SC}$ increases from 58.85 to 59.31 mA/cm$^2$ when the temperature is raised from 273 to 500 K. On the contrary, the $V_{OC}$ drops from 0.96 to 0.75 V for the same temperature range, which can be explained in terms of the increment of the recombination current. The *FF* also decreases at higher temperatures, causing the PCE to change from 45.64 to 29.89% as the temperature increases from 273 to 500 K.

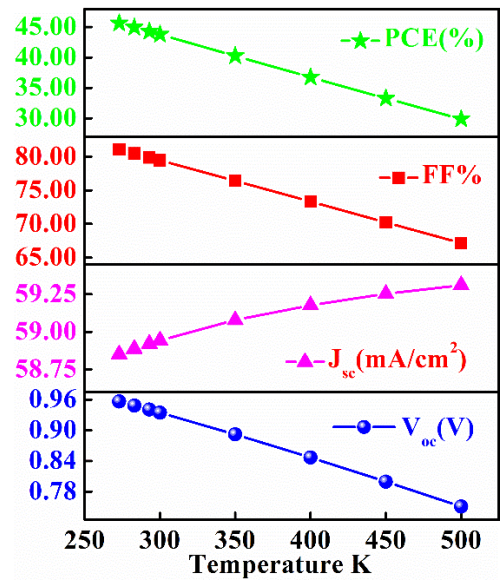

**Fig.6**. PV parameters with respect to operating temperature.

*3.6 Single vs dual heterojunction cells*

Figure 7 (a) and (b) show the simulated J-V and QE curves of CuSbSe$_2$-based single and dual heterojunction solar cells, respectively. The *Jsc* in the dual- heterojunction is increased by more than 11.5 mA/cm-$^2$ compared to the single heterojunction cell. This is because of the CGS BSF layer which also performs as bottom absorber layer and absorbs longer wavelength photons through TSA two-step photo upconversion process. The CGS layer also increases the built-in potential of the CuSbSe$_2$/CGS interface, leading to a higher $V_{OC}$. Figure 7(b) shows that the QE for dual heterojunction is around 51% towards longer wavelengths beyond 2000 nm. On the other hand, single heterojunctions, without the CGS, layer, yields 0% QE for the same wavelength ranges beyond 2000 nm.

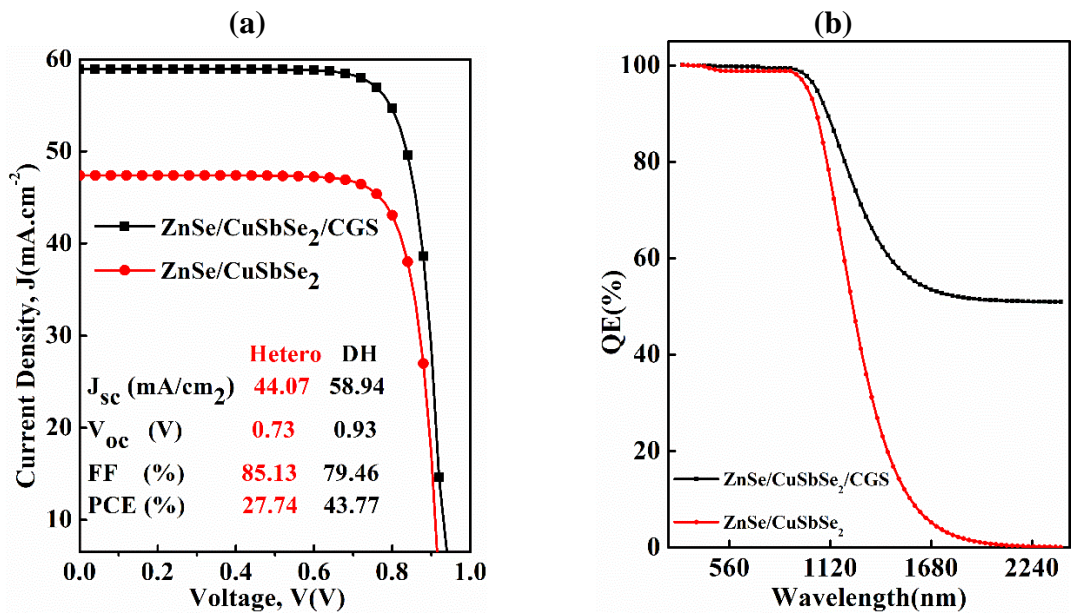

**Fig.7**. Simulated J-V and QE Cell performance for CuSbSe$_2$-based single and dual heterojunction solar cells.

## 4. Conclusion

In this work, we used numerical simulations to demonstrate a high-performance *n*-ZnSe/*p*-CuSbSe$_2$/$p^+$-CGS dual heterojunction solar cell device. Results confirm that the inclusion of CGS as back surface field (BSF) layer in the photovoltaic structure significantly improves the PCE through a (TSA) two-step photon upconversion process. The highest efficiency obtained is 43.77 %, when $V_{oc}$ = 0.9345 V, $J_{sc}$ = 58.943 mA/cm$^2$ and FF = 79.46 %, respectively. The findings in this work pave the way for future experimental studies on CuSbSe$_2$ and CGS compounds as promising absorber and BSF layers, respectively, in next generation thin film solar cells.

## Acknowledgements

The authors highly appreciate Dr. Marc Burgelman, University of Gent, Belgium, for providing SCAPS simulation software.

**Corresponding Authors:**

*E-mail: jak_apee@ru.ac.bd (Jaker Hossain).

**NOTES:** The authors declare no competing financial interest.